\documentclass[runningheads]{llncs}
\usepackage{amsmath,amsfonts}
\usepackage{algorithmic}
\usepackage{hyperref}
\usepackage{caption}
\usepackage{graphicx}
\usepackage{algorithm}
\usepackage{array}
\usepackage{color,soul}
\usepackage{multirow}
\usepackage[svgnames]{xcolor}
\usepackage{tikz}
\usetikzlibrary{calc}

\usepackage{array}        
\usepackage{booktabs}     
\usepackage{amsmath}      
\usepackage{xcolor}       

\usepackage[caption=false]{subfig}
\usepackage{amssymb}
\usepackage{pifont}
\usepackage{textcomp}
\usepackage{stfloats}
\usepackage{url}
\usepackage{booktabs}
\usepackage{hhline}
\usepackage{verbatim}
\usepackage{graphicx}
\usepackage{cite}
\usepackage{diagbox}
\usepackage{tabularx,ragged2e}
\usepackage{orcidlink}
\usepackage{lettrine}
\usepackage[T1]{fontenc}
\usepackage{graphicx}
\usepackage{enumitem}
\setlist[itemize]{label=$\bullet$}
\usepackage{color}

\begin{document}
\title{Approximate Signed Multiplier with Sign-Focused Compressor for Edge Detection Applications}
\titlerunning{Approximate Signed Multiplier for Edge Detection Applications}
%

\author{
L.Hemanth Krishna\inst{1} ~\orcidlink{0000-0002-7737-6685} \and
Srinivasu Bodapati\inst{1} ~\orcidlink{0000-0003-0974-8245} \and
Sreehari Veeramachaneni\inst{2}~\orcidlink{0000-0001-7744-4580} \and
BhaskaraRao Jammu\inst{3}~\orcidlink{0000-0001-5878-7780} \and
Noor Mahammad Sk\inst{4}~\orcidlink{0000-0003-4708-4769}
}
\authorrunning{L. Hemanth Krishna et al.}

\institute{Indian Institute of Technology, Mandi,\and 
Sri Sivasubramaniya Nadar College of Engineering, Chennai, \and
GVP College of Engineering, Visakhapatnam,  \and
Indian Institute of Information Technology Design \& Manufacturing, Kancheepuram,
}
\maketitle              
\begin{abstract}

This paper presents an approximate signed multiplier architecture that incorporates a sign-focused compressor, specifically designed for edge detection applications in machine learning and signal processing.
The multiplier incorporates two types of sign-focus compressors
$A + B + C + 1$ and $A + B + C + D + 1$. 
Both exact and approximate compressor designs are utilized, 
with a focus on efficiently handling constant value ‘1’ and 
negative partial products, 
which frequently appear in the partial product matrices of signed multipliers. 
To further enhance efficiency, the lower $N-1$ columns of the partial product 
matrix is truncated, followed by an error compensation mechanism. 
Experimental results show that the proposed 8-bit approximate multiplier achieves 
a 29.21\% reduction in power delay product (pdp) and a 14.39\% 
reduction in power compared to its best of existing multiplier. 
The proposed multiplier is integrated into a 
custom convolution layer and perform  edge detection, 
demonstrating its practical utility in real-world applications.
\end{abstract}
\section{Introduction}
\lettrine[lines=2, lhang=0, loversize=0.1]{\textbf{R}}{apid} growth of machine learning and signal processing applications has heightened 
the demand for energy-efficient arithmetic units, 
particularly multipliers, which are core components in many computing systems~\cite{han2013approximate}. 
Convolutional Neural Networks (CNNs), widely used in image processing tasks such as edge detection, 
involve a large number of multiply-and-accumulate operations. 
Consequently, designing low-power and high-performance multipliers is 
critical for reducing the computational burden 
and power consumption of such systems~\cite{venkataramani2015approximate}.

Approximate computing has emerged as a promising design paradigm for applications 
that can tolerate some loss in numerical accuracy in exchange 
for improved performance metrics such as 
area, power, and delay~\cite{mittal2016survey}.
In general, the Booth algorithm~\cite{Booth1975} and the 
Baugh-Wooley algorithm~\cite{baugh1974} 
are the two most widely used techniques for 
performing signed multiplication in digital systems. 
Among them, the Baugh-Wooley algorithm offers a 
simpler and more structured approach for 
generating the Partial Product Matrix (PPM), 
as illustrated in Figure~\ref{fig:ppm_baugh}. 
Due to the regularity and predictability of the partial product generation, 
the Baugh-Wooley method is particularly well-suited for applications 
involving \textit{approximate computing}, where controlled trade-offs 
between accuracy and efficiency are desirable.\\

\noindent\textbf{The main contributions of this paper are as follows:}
\begin{itemize}
    \item This paper proposes two sign-focused compressor architectures, $A+B+C+1$ and $A+B+C+D+1$, to effectively handle constant '1' values and negative partial products, which are prevalent in signed multiplier partial product matrices.
    \item A truncation strategy is applied to the lower \(N-1\) columns of the partial product matrix, followed by an error compensation technique to balance performance and accuracy.
    \item The proposed approximate 8-bit multiplier achieves significant hardware savings, including 14.39\% power reduction and 29.21\% power-delay product reduction, compared to the best of existing multipliers~\cite{Du2022}.
    \item The multiplier is successfully integrated into a custom convolution layer, and its effectiveness is demonstrated through an edge detection task.
\end{itemize}

The rest of the paper is organized as follows. 
Section~\ref{sec:baugh_wooley} discusses the implementation of signed multiplication 
using the Baugh-Wooley algorithm and introduces exact and approximate signed compressors. 
Section~\ref{sec:proposed_multiplier} presents the architecture of the proposed approximate signed multiplier 
using sign-focused compressors. 
Section~\ref{sec:edge_detection} demonstrates the application of the proposed design in convolution-based edge detection. 
Simulation results, including error and hardware analysis, are detailed in Section~\ref{sec:results}. 
Finally, Section~\ref{sec:conclusion} concludes the paper.

\section{Implementation of Signed Multiplication Using Baugh-Wooley Algorithm}
\label{sec:baugh_wooley}

Signed multipliers play a crucial role in the hardware implementation of deep learning accelerators, 
particularly in convolutional neural networks (CNNs) and deep neural networks (DNNs).
In general, the Baugh-Wooley algorithm~\cite{baugh1974} is the most widely used technique for performing signed multiplication in digital systems. 

Let \( A \) and \( B \) be two signed \( N \)-bit numbers represented in 2's complement format, as defined in Equation~\ref{eq:sign}.
\begin{eqnarray}
A &=& -a_{N-1} \cdot 2^{N-1} + \sum_{i=0}^{N-2} a_i \cdot 2^i \notag \\
B &=& -b_{N-1} \cdot 2^{N-1} + \sum_{j=0}^{N-2} b_j \cdot 2^j \label{eq:sign}
\end{eqnarray}

The product \( A \times B \) is expressed in Equation~\ref{eq:ab}:
\begin{eqnarray}
A \times B &=& \left( -a_{N-1} \cdot 2^{N-1} + \sum_{i=0}^{N-2} a_i \cdot 2^i \right) \nonumber \\
&&\times \left( -b_{N-1} \cdot 2^{N-1} + \sum_{j=0}^{N-2} b_j \cdot 2^j \right) \label{eq:ab}
\end{eqnarray}

The multiplication in Equation~\ref{eq:ab} can be expanded as shown in Equation~\ref{eq:expand}, where each term corresponds to a specific group of partial products, categorized by the involvement of sign or non-sign bits.

\begin{align}
A \times B =\;
&\sum_{i=0}^{N-2} \sum_{j=0}^{N-2} a_i b_j \cdot 2^{i+j}
\;-\;
\sum_{i=0}^{N-2} a_i b_{N-1} \cdot 2^{i+N-1} \nonumber \\
&\;-\; 
\sum_{j=0}^{N-2} a_{N-1} b_j \cdot 2^{j+N-1}
\;+\;
a_{N-1} b_{N-1} \cdot 2^{2N-2} \label{eq:expand}
\end{align}

\begin{figure}
	\centerline{\includegraphics[scale=0.54]{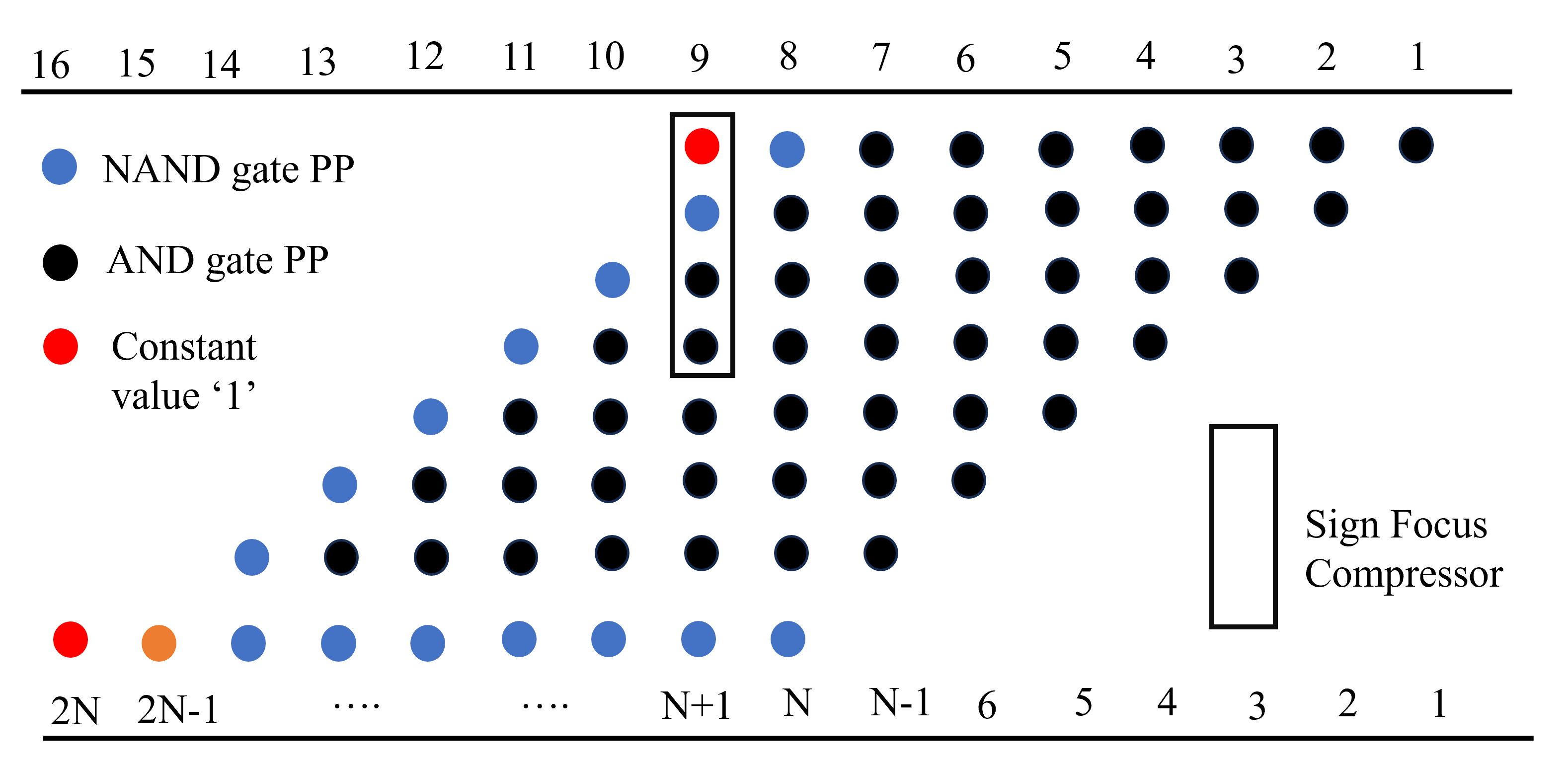}}
        \caption{Baugh-Wooley algorithm for signed $8 \times 8$ multiplication \cite{baugh1974}. }
        \label{fig:ppm_baugh}
\end{figure}

In Equation~\ref{eq:expand}, the positive partial products are generated using AND gates, 
represented by black dots in Fig.~\ref{fig:ppm_baugh}, 
while the negative partial products are generated using NAND gates, represented by blue dots in the same figure.
In a signed Baugh-Wooley multiplier, constant values of 1 are added to the 
\( (N+1)^{\text{th}} \) and \( 2N^{\text{th}} \) columns. 
Table~\ref{Tab:baug} illustrates this behavior clearly with an example for \( N = 4 \). 
It demonstrates the generation of partial products, 
the application of two’s complement for handling negative partial products, 
and how this process results in constant '1's appearing in the \( 2^4 \) and \( 2^7 \) positions, 
corresponding to the \( (N+1)^{\text{th}} \) and \( 2N^{\text{th}} \) columns respectively.

\begin{table}[h]
\centering
\caption{Example of Baugh-Wooley Multiplication for \(N = 4\)}
\label{Tab:baug}
\setlength{\tabcolsep}{4pt}  
\renewcommand{\arraystretch}{1.2}  
\begin{tabular}{c*{8}{>{\centering\arraybackslash}p{1.2cm}}}
\toprule
\textbf{Sign} & \(2^7\) & \(2^6\) & \(2^5\) & \(2^4\) & \(2^3\) & \(2^2\) & \(2^1\) & \(2^0\) \\
\midrule
+ & 0 & \(a_3b_3\) & 0 & 0 & 0 & \(a_0b_2\) & \(a_0b_1\) & \(a_0b_0\) \\
+ & 0 & 0 & 0 & 0 & \(a_1b_2\) & \(a_1b_1\) & \(a_1b_0\) & ~ \\
+ & 0 & 0 & 0 & \(a_2b_2\) & \(a_2b_1\) & \(a_2b_0\) & ~ & ~ \\
$-$ & 0 & 0 & \(a_2b_3\) & \(a_1b_3\) & \(a_0b_3\) & ~ & ~ & ~ \\
$-$ & 0 & 0 & \(a_3b_2\) & \(a_3b_1\) & \(a_3b_0\) & ~ & ~ & ~ \\
\midrule
\multicolumn{9}{c}{\textit{The last two rows are negative and are converted using two’s complement.}} \\
\midrule
+ & 0 & \(a_3b_3\) & 0 & 0 & 0 & \(a_0b_2\) & \(a_0b_1\) & \(a_0b_0\) \\ 
+ & 0 & 0 & 0 & 0 & \(a_1b_2\) & \(a_1b_1\) & \(a_1b_0\) & ~ \\ 
+ & 0 & 0 & 0 & \(a_2b_2\) & \(a_2b_1\) & \(a_2b_0\) & ~ & ~ \\ 
+ & \textcolor{red}{1} & \textcolor{red}{1} & \(\overline{a_2b_3}\) & \(\overline{a_1b_3}\) & \(\overline{a_0b_3}\) & ~ & ~ & ~ \\
\multicolumn{6}{r}{\textcolor{blue}{+1}} \\
+ & \textcolor{red}{1} & \textcolor{red}{1} & \(\overline{a_3b_2}\) & \(\overline{a_3b_1}\) & \(\overline{a_3b_0}\) & ~ & ~ & ~ \\
\multicolumn{6}{r}{\textcolor{blue}{+1}} \\
\midrule
\multicolumn{9}{c}{\textit{Final addition results in ‘1’ at positions \(2^4 = 2^N\) and \(2^7 = 2^{2N-1}\).}} \\
\midrule
+ & \textcolor{red}{1} & \(a_3b_3\) & \(\overline{a_3b_2}\) & \(\overline{a_3b_1}\) & \(\overline{a_3b_0}\) & & & \\
+ & & & \(\overline{a_2b_3}\) & \(a_2b_2\) & \(a_2b_1\) & \(a_2b_0\) & & \\
+ & & & & \(\overline{a_1b_3}\) & \(a_1b_2\) & \(a_1b_1\) & \(a_1b_0\) & \\
+ & & & & \textcolor{red}{1} & \(\overline{a_0b_3}\) & \(a_0b_2\) & \(a_0b_1\) & \(a_0b_0\) \\
\bottomrule
\end{tabular}
\end{table}

\subsection{Existing Signed Exact Compressors}
The signed exact compressor proposed by the authors in \cite{Du2022} performs the summation of three input bits along with a constant value of 1, i.e., \( A + B + C + 1 \). It generates three output bits \textit{cout}, \textit{carry}, and \textit{sum}. 
Unlike typical compressors, this design does not reduce any partial product bits during the compression stage.
Furthermore, to implement this exact compressor, 
additional hardware components such as XOR gates are required, 
which increases the overall circuit complexity. 

\begin{figure}[h]
    \centering
    \includegraphics[width=0.98\linewidth]{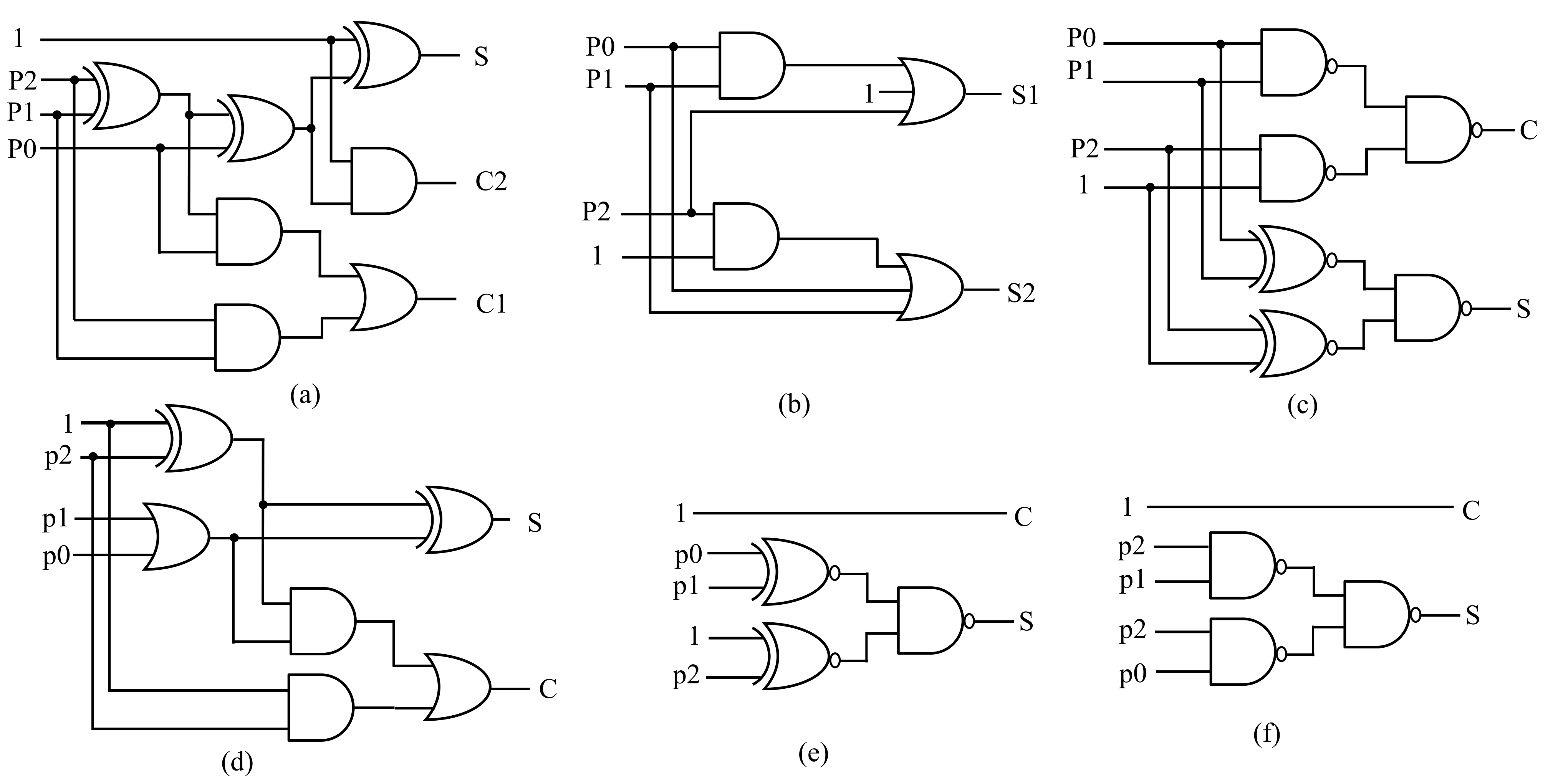} 
    \caption{Schematic comparison of compressor (a) Exact Compressor \cite{Du2022} (b) Approximate compressor \cite{Esposito2018} (c) Approximate compressor \cite{Guo2019} (d) Approximate compressor \cite{Strollo2020}  (e) Approximate compressor \cite{Akbari2017} (f) Approximate compressor \cite{Laimin}.}
\label{fig:exst_AC}
\end{figure}

In~\cite{Laimin}, an approximate sign-focused compressor was proposed for use in signed multipliers, 
specifically targeting positions in the partial product matrix (PPM) 
where a constant value of 1 is added, as illustrated in Fig.~\ref{fig:ppm_baugh}. 
This design introduces errors in three input combinations, resulting in an error
probability (\(P_E\)) of \(18/64\) and a mean error ($E_{mean}$) of \(-0.2812\).
The design generates the \textit{Sum} output while maintaining the \textit{Carry} output
as a constant value of 1. Its hardware implementation is shown in Fig.~\ref{fig:exst_AC}.
\begin{align}
Err_i &= S - S_{\text{APP}} \nonumber \\
P_E &= \sum_{i} P(\text{Err}_i) \nonumber \\
E_{\text{mean}} &= \sum_{i} P(\text{Err}_i) \cdot \text{Err}_i
\label{eq:mep}
\end{align}
Similarly, the compressor presented in~\cite{Esposito2018} exhibits inaccuracies for input combinations involving the summation of three or four logical ones, leading to a total of four error cases. This design yields an error probability of \(22/64\) and a corresponding mean error probability of \(25/64\), as derived using Equation~(\ref{eq:mep}). In comparison, the compressor proposed in~\cite{Guo2019} introduces errors in three specific input combinations, resulting in a mean error ($E_{mean}$)of \(0.1875\). Additionally, the design introduced in~\cite{Strollo2020,Krishna2024}, which employs a stacking logic mechanism, incurs errors in four input combinations and demonstrates a notably higher mean error ($E_{mean}$) of \(0.75\).
The circuit architectures of all these existing designs are illustrated in Fig.~\ref{fig:exst_AC}, and their corresponding truth tables, detailing the specific error patterns and outputs, are provided in Table~\ref{tab:AC}.
In this table, input \( A \) represents the negative partial product, 
while inputs \( B \) and \( C \) correspond to the positive partial products. The corresponding error probability and mean error for each design are  also included

\begin{table*}[!ht]
\centering
\caption{Truth Table and Error Comparison of Sign-Focused A+B+C+1 Compressors}
\label{tab:AC}
\resizebox{\textwidth}{!}{%
\begin{tabular}{|c|c|c|c|c|c|c|c|c|c|c|c|c|c|c|c|c|c|c|c|c|}
\hline
\multicolumn{5}{|c|}{Input Cases} & Exact & \multicolumn{2}{c|}{AC1~\cite{Esposito2018}} & \multicolumn{2}{c|}{AC2~\cite{Guo2019}} & \multicolumn{2}{c|}{AC3~\cite{Strollo2020}} & \multicolumn{2}{c|}{AC4~\cite{Laimin}} & \multicolumn{2}{c|}{AC5~\cite{Du2022}} & \multicolumn{4}{c|}{Proposed} \\ \hline
Const & P2 & P1 & P0 & $P(Err)$ & $S_{\text{exact}}$ & $S_{\text{aprx}}$ & Err & $S_{\text{aprx}}$ & Err & $S_{\text{aprx}}$ & Err & $S_{\text{aprx}}$ & Err & $S_{\text{aprx}}$ & Err & Carry & Sum & $S_{\text{aprx}}$ & Err \\ \hline

1 & 0 & 0 & 0 & $9/64$  & 1 & 1 & 0 & 1 & 0 & 1 & 0 & 3 & \textcolor{red}{+2} & 2 & \textcolor{red}{+1} & 0 & 1 & 1 & 0 \\ \hline
1 & 0 & 0 & 1 & $3/64$  & 2 & 2 & 0 & 1 & \textcolor{red}{-1} & 2 & 0 & 3 & \textcolor{red}{1} & 2 & 0 & 1 & 1 & 3 & \textcolor{red}{+1} \\ \hline
1 & 0 & 1 & 0 & $3/64$  & 2 & 2 & 0 & 1 & \textcolor{red}{-1} & 2 & 0 & 3 & \textcolor{red}{1} & 2 & 0 & 1 & 1 & 3 & \textcolor{red}{+1} \\ \hline
1 & 0 & 1 & 1 & $1/64$  & 3 & 2 & \textcolor{red}{-1} & 3 & 0 & 3 & 0 & 3 & 0 & 2 & \textcolor{red}{-1} & 1 & 1 & 3 & 0 \\ \hline
1 & 1 & 0 & 0 & $27/64$ & 2 & 2 & 0 & 2 & 0 & 1 & \textcolor{red}{-1} & 2 & 0 & 2 & 0 & 1 & 0 & 2 & 0 \\ \hline
1 & 1 & 0 & 1 & $9/64$  & 3 & 2 & \textcolor{red}{-1} & 3 & 0 & 2 & \textcolor{red}{-1} & 3 & 0 & 3 & 0 & 1 & 1 & 3 & 0 \\ \hline
1 & 1 & 1 & 0 & $9/64$  & 3 & 2 & \textcolor{red}{-1} & 3 & 0 & 2 & \textcolor{red}{-1} & 3 & 0 & 3 & 0 & 1 & 1 & 3 & 0 \\ \hline
1 & 1 & 1 & 1 & $3/64$  & 4 & 2 & \textcolor{red}{-2} & 2 & \textcolor{red}{-2}& 3 & \textcolor{red}{-1} & 2 & \textcolor{red}{-2} & 3 & \textcolor{red}{1} & 1 & 1 & 3 & \textcolor{red}{-1} \\ \hline
\multicolumn{5}{|c|}{Error Probability ($P_E$)} & 0 & \multicolumn{2}{c|}{0.3437} & \multicolumn{2}{c|}{0.1406} & \multicolumn{2}{c|}{0.7500} & \multicolumn{2}{c|}{0.2813} & \multicolumn{2}{c|}{0.2031} & \multicolumn{4}{c|}{\textcolor{purple}{0.0140}} \\ \hline
\multicolumn{5}{|c|}{Mean Erro ($E_{mean}$)} & 0 & \multicolumn{2}{c|}{0.3906} & \multicolumn{2}{c|}{0.1875} & \multicolumn{2}{c|}{0.7500} & \multicolumn{2}{c|}{-0.2813} & \multicolumn{2}{c|}{-0.0781} & \multicolumn{4}{c|}{\textcolor{purple}{-0.0468}} \\ \hline
\end{tabular}}
\end{table*}

\section{Proposed Compressor-Based Approximate Signed Multiplier Design}
\label{sec:proposed_multiplier}
This section presents the architecture of the proposed signed approximate multiplier,
which is developed using a sign-focused compressor. 
The discussion begins with the proposed sign-focused compressor,
followed by a detailed explanation of the proposed approximate multiplier architecture.

\subsection{Proposed Sign-Focus Compressors}
In this work, we propose both exact and approximate sign-focus compressor designs. 
Specifically, the proposed exact sign-focused compressor performs the summation of four inputs
along with a logic '1' (i.e.,  $A + B + C + D + 1$ ), and generates three output bits \textit{cout}, \textit{carry}, and \textit{sum}. 
Unlike the existing exact compressor in \cite{Du2022}, which does not reduce any partial product,
the proposed design effectively reduces one partial product at a time, contributing to improved 
compression efficiency in the multiplier architecture.

\begin{figure}[h!]
\centering
\includegraphics[scale=0.5]{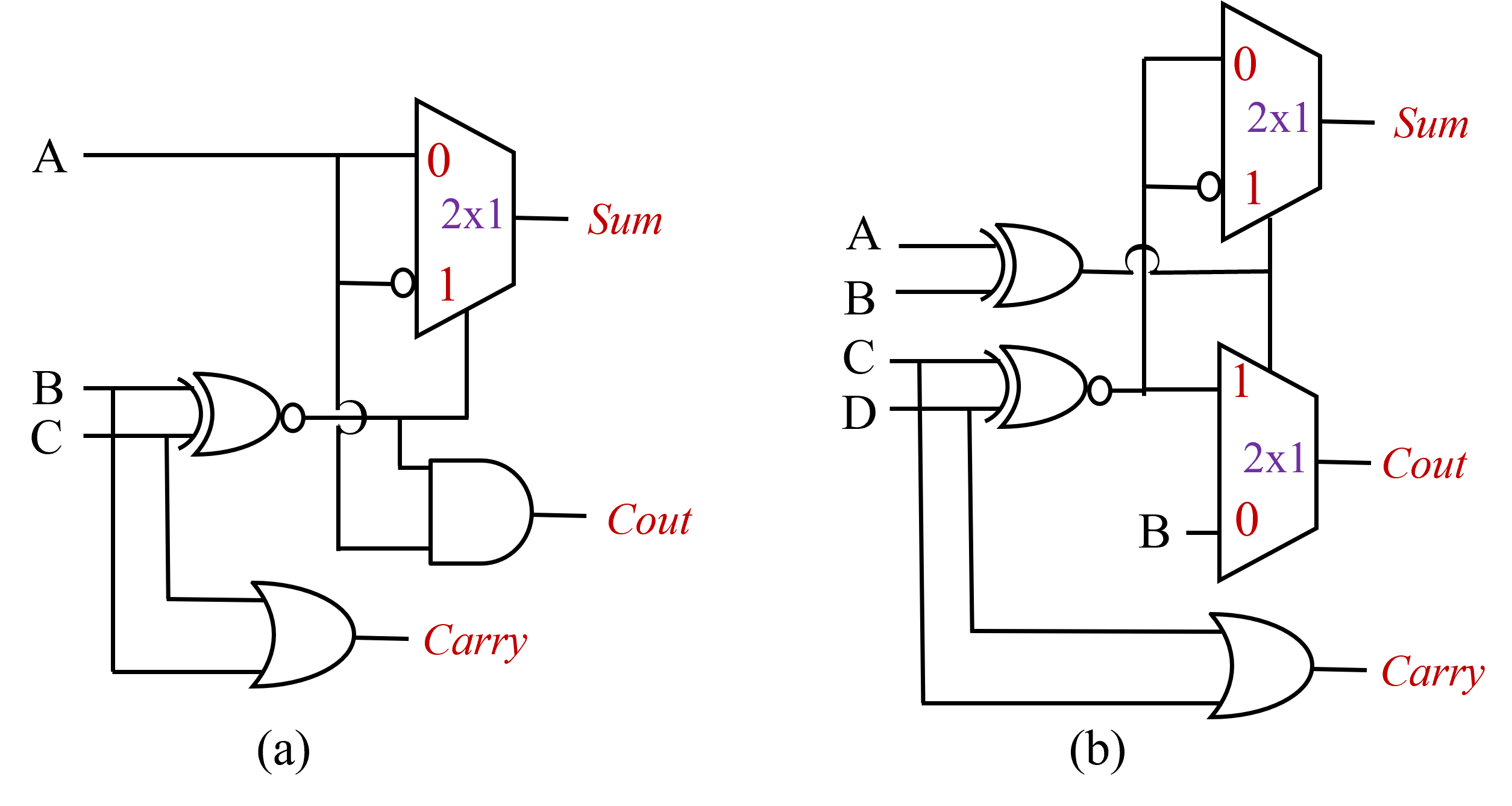}
\caption{Proposed exact sign-focus compressors: (a) $A + B + C + 1$, (b) $A + B + C + D + 1$}
\label{fig:exact_prp}
\end{figure}

Two exact compressor designs are proposed 
in this work, such as \( A + B + C + 1 \), and \( A + B + C + D + 1 \). 
The corresponding circuit implementations of these compressors are illustrated in Figure~\ref{fig:exact_prp} (a) and  (b), respectively.
These compressors are primarily utilized in the Center Significant Part (CSP) region of the partial product matrix 
to preserve accuracy in significant bit positions.
While designing the proposed approximate sign-focus compressors, the error is intentionally introduced 
in the sum output rather than the carry output to reduce the overall error. 
The introduced error in the sum output follows a probabilistic distribution, with errors being 
introduced at input combinations with lower probabilities. The proposed  \( A + B + C + 1 \) compressor design achieves a lower error probability ($P_E$)
and a reduced mean error (\( M_{\text{error}} \)) as compared to other existing compressor designs, as shown in Table \ref{tab:AC}.
\begin{figure}[ht]
    \centering
    \includegraphics[scale=0.5]{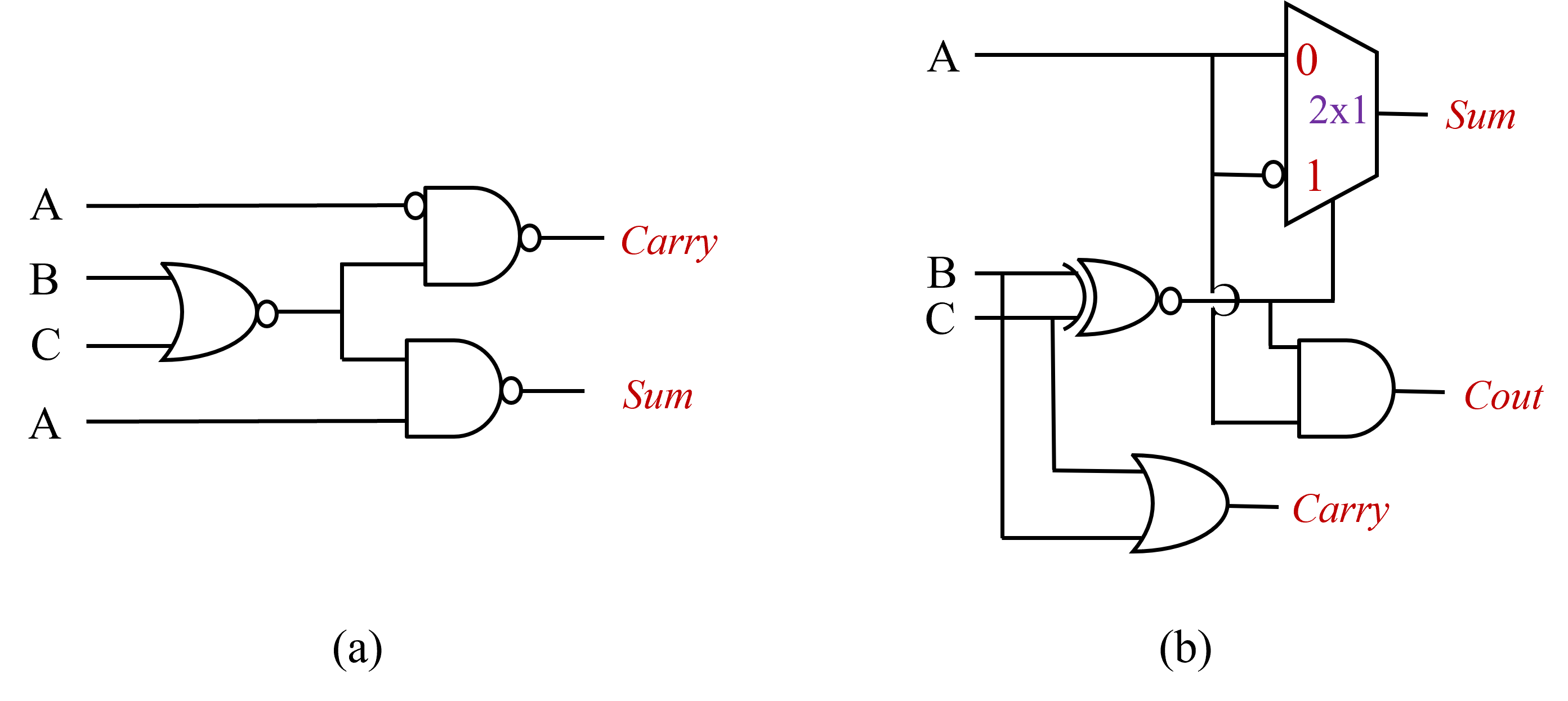}
    \caption{Proposed approximate sign-focus compressor designs: (a) $A + B + C + 1$, (b) $A + B + C + D + 1$}
    \label{fig:aprx2}
\end{figure}
\begin{table}
	\centering
	\caption{Truth table of  \( A + B + C + D + 1 \) compressor
                 where A is the negative partial product.}
	\label{tab:sign_4I}
	\begin{tabular}{|c|c|c|c|c|c|c|c|c|c|c|c|}\hline
        \multicolumn{6}{|c|}{Inputs} & \multicolumn{3}{c|}{Exact} & \multicolumn{3}{c|}{Approximate}\\ \hline
		Const&A & B & C & D & P(E) & Cout & Carry & Sum  & Carry & Sum &ED\\\hline
		1 & 0 & 0 & 0 & 0 & 27/256& 0 & 0 & 1 & 0 & 1 & 0\\
		\hline
		1 &0 & 0 & 0 & 1 & 9/256 & 1 & 0 & 0 & 1 & 0 & 0\\
		\hline
		1 &0 & 0 & 1 & 0 & 9/256 & 0 & 1 & 0 & 1 & 0 & 0\\
		\hline
		1 &0 & 0 & 1 & 1 & 3/256 & 0 & 1 & 1 & 1 & 0 & \textcolor{red}{1} \\
		\hline
		1 &0 & 1 & 0 & 0 & 9/256 & 0 & 1 & 0 & 1 & 0 & 0\\
		\hline
		1 &0 & 1 & 0 & 1 & 3/256 & 0 & 1 & 1 & 1 & 1 & 0\\
		\hline
		1 &0 & 1 & 1 & 0 & 3/256 & 0 & 1 & 1 & 1 & 1 & 0\\
		\hline
		1 &0 & 1 & 1 & 1 & 1/256 & 1 & 1 & 0 & 1& 1 & \textcolor{red}{1} \\
		\hline
		1 &1 & 0 & 0 & 0 & 81/256 & 1 & 0 & 0 & 1 & 0 & 0\\
		\hline
		1 &1 & 0 & 0 & 1 & 27/256 & 1 & 0 & 1 & 1 & 1 & 0\\
		\hline
		1 &1 & 0 & 1 & 0 & 27/256 & 0 & 1 & 1 & 1 & 1 & 0\\
		\hline
		1 &1 & 0 & 1 & 1 & 9/256 & 1 & 1 & 0 & 1 & 1 & \textcolor{red}{1} \\
		\hline
		1 &1 & 1 & 0 & 0 & 27/256& 0 & 1 & 1 & 1 & 1 & 0\\
		\hline
		1 &1 & 1 & 0 & 1 & 9/256 & 1 & 1 & 0 & 1 & 1 & \textcolor{red}{1} \\
		\hline
		1 &1 & 1 & 1 & 0 & 9/256 & 1 & 1 & 0 & 1 & 1 & \textcolor{red}{1} \\
		\hline
		1 &1 & 1 & 1 & 1 & 3/256 & 1 & 1 & 1 & 1 & 1 & \textcolor{red}{2} \\
		\hline
	\end{tabular}
\end{table}
Table~\ref{tab:sign_4I} presents the truth table of the proposed approximate signed-focus compressor 
of $A + B + C + D + 1$. This compressor computes the summation of three positive 
partial products and one negative partial product, along with a constant logic value of 1. The probability 
analysis of the proposed design is derived from the combined probabilities of the AND and NAND gates. 
In that table, the input \( A \) represents the negative partial product (realized using a NAND gate), 
while \( B \), \( C \), and \( D \) represent the positive partial products (realized using AND gates). 
The corresponding circuit implementation of the proposed approximate compressor is shown in Fig.~\ref{fig:aprx2}(b).

\subsection{Approximate Signed Multiplier Architecture}
\begin{figure}[h]
    \centering
    \includegraphics[width=0.88\linewidth]{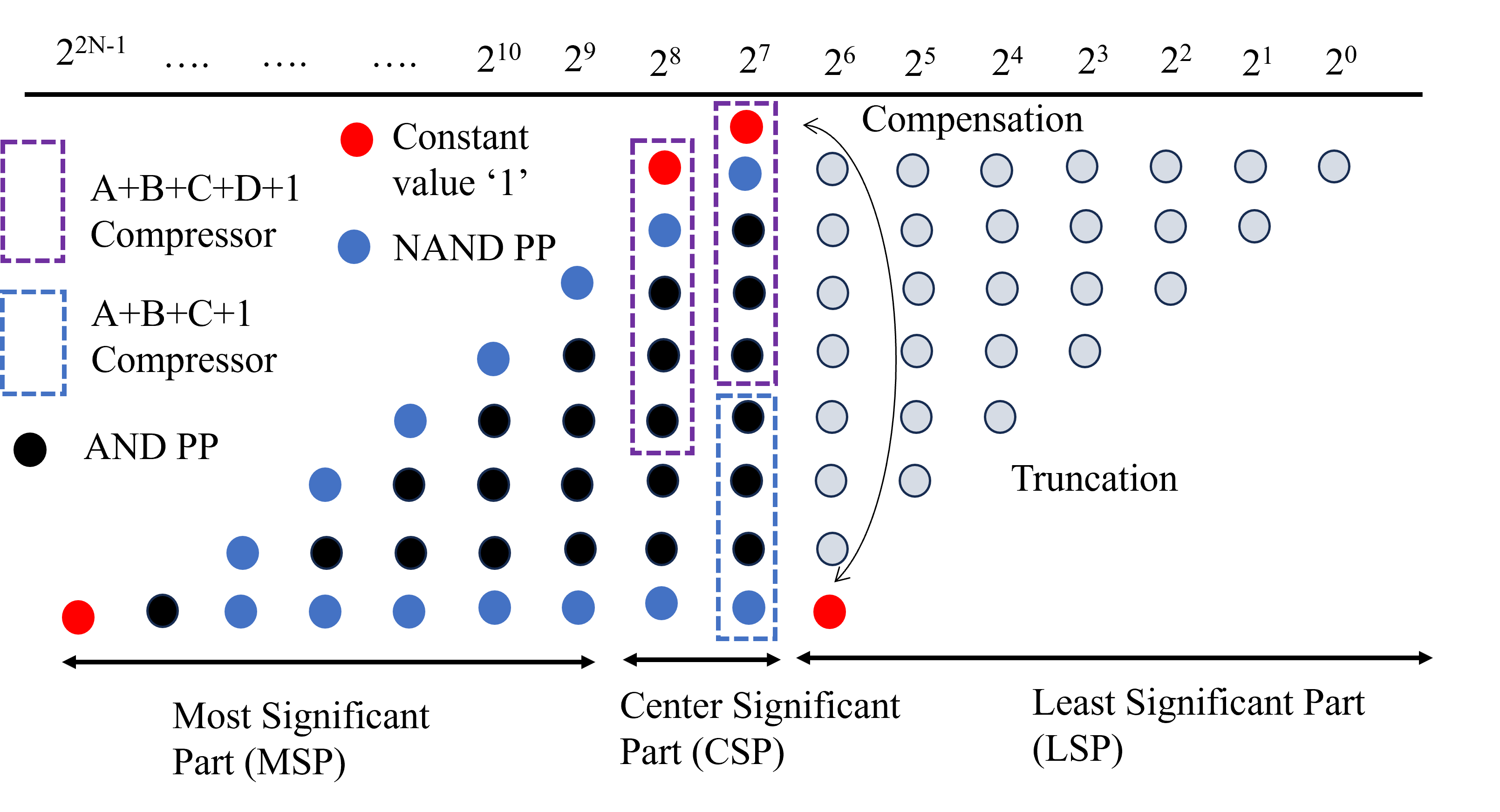}
    \caption{Partial product column partitioning in the proposed approximate signed multiplier architecture}
\label{fig:prp_mul_1}
\end{figure}

In an $N \times N$  signed multiplier, the partial product matrix consists of \( 2N \) columns. 
For analysis purposes, these columns are divided into three distinct regions, such as the Least Significant Part (LSP), 
the Center Significant Part (CSP), and the Most Significant Part (MSP). 
Specifically, the LSP comprises the first \( n - 1 \) columns, 
the CSP includes two central columns, namely columns \( n \) and \( n + 1 \), 
and the MSP consists of the remaining columns. 
This partitioning is illustrated in the proposed multiplier architecture shown in Fig.~\ref{fig:prp_mul_1}.

Since the sign-focused compressor requires a constant logic value of 1 for proper correction, the CSP region of the architecture (as illustrated in Fig.~\ref{fig:prp_mul_1}) includes only a single constant logic 1 input. 
To address this, negative partial products are converted to a constant logic 1. Given that the probability of a NAND gate producing an output of 1 is \(\frac{3}{4}\), this probabilistic behavior is leveraged to simplify the design. 
One NAND-based partial product is replaced with a constant logic 1 and incorporated into the sign-focused compressor at the \(2^N\) column. Additionally, two constant logic 1's are added at the \(N\) and \(N-1\) columns as part of the error compensation technique, which will be explained in the following subsection.
\subsection{Error Compensation}
In the proposed approximate multiplier architecture, the least significant part (LSP) 
comprising the lower \( N-1 \) columns of the partial product matrix is truncated to 
reduce area and power consumption. However, this approximation introduces 
a deterministic error due to the omission of partial product contributions 
from these columns, which, though small in individual weight, 
can cumulatively impact the accuracy of the final result.

To address this, an efficient error compensation technique is introduced. 
This method relies on a probabilistic estimation of the missing partial product contributions. 
Each partial product bit in the LSP is formed by an 
AND gate operation, i.e., \( P_{ij} = x_i \cdot y_j \). 
Assuming the input bits \( x_i \) and \( y_j \) are independent and uniformly 
distributed, the probability that \( P_{ij} = 1 \) is \( \frac{1}{4} \). 
Using this, the expected accumulated value of the truncated partial products can be expressed as Equation~(\ref{eq:trunc_error})
\begin{equation}
T_T = \sum_{q=0}^{N-2} \left( \frac{1}{4} \right) \cdot (q + 1) \cdot 2^q
\label{eq:trunc_error}
\end{equation}

where \( q \) denotes the column index, \( (q+1) \) represents the number of partial products in that column, and \( 2^q \) is the weight of the corresponding bit position.
To compensate for this estimated error without significantly increasing complexity or power, a constant logic ‘1’ is added to both the \( N \)th and \( (N-1) \)th columns of the partial product matrix. 
This compensation method effectively improves the output accuracy of the approximate multiplier while preserving its key benefits of reduced delay, area, and energy consumption.
\begin{figure}[h]
    \centering
    \includegraphics[width=0.88\linewidth]{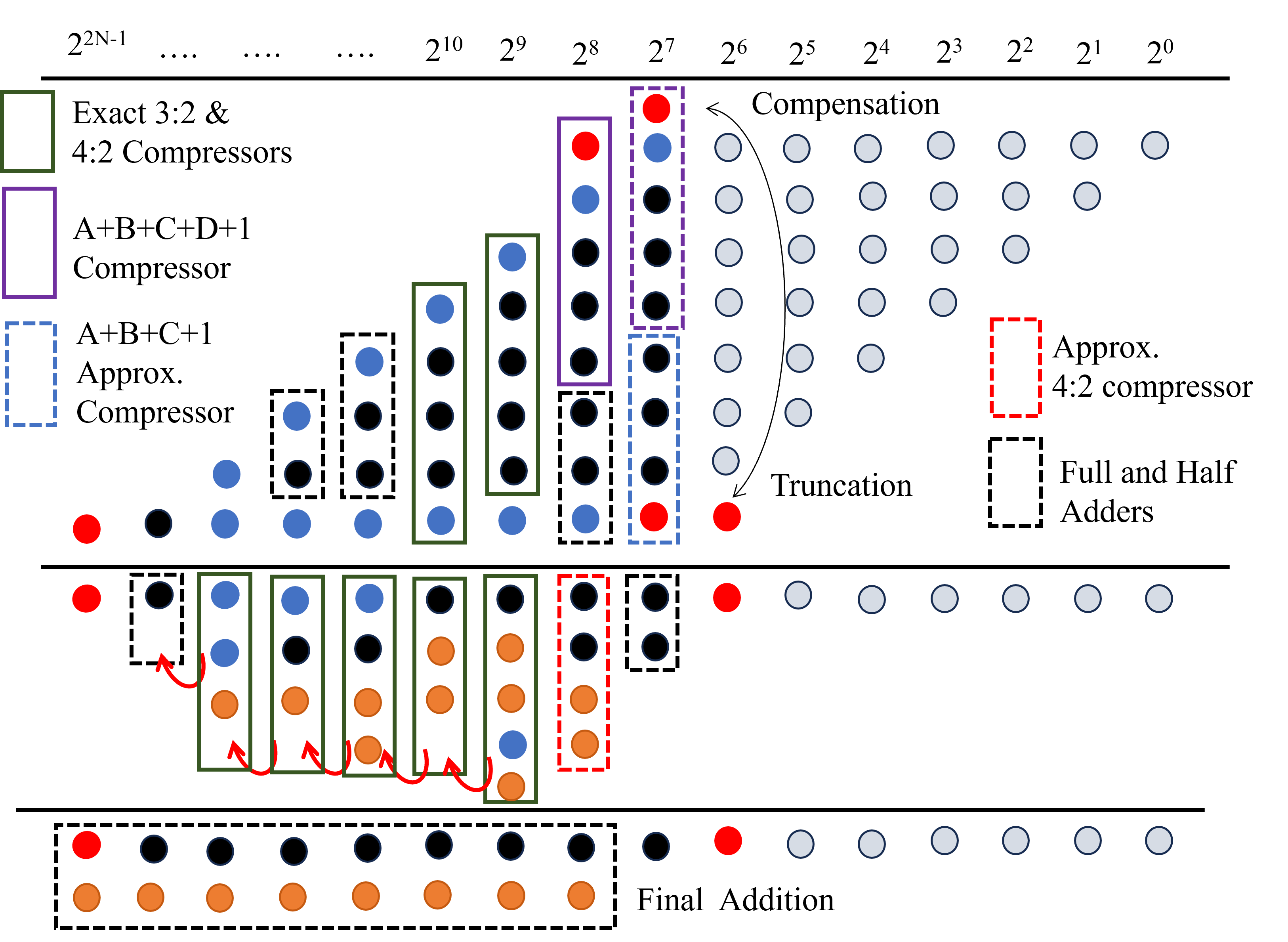} 
    \caption{ Proposed Approximate Signed Multiplier Architecture}
\label{fig:prp_mul}
\end{figure}

Fig.~\ref{fig:prp_mul} illustrates the proposed signed approximate multiplier architecture. 
The design incorporates a total of three sign-focused compressors 
within the Center Significant Part (CSP) to improve efficiency. 
In the Most Significant Part (MSP), a combination of adders and compressors as presented in~\cite{hemanth2025} are used to add the partial products to ensure accurate computation.
After the partial product reduction stage, the final summation is performed using an $N$-bit carry-save adder (CSA) structure,
which completes the multiplication in three stages, including the final addition.

The hardware complexity of the proposed signed multiplier design uses seven exact compressors~\cite{hemanth2025} designs and one approximate compressor~\cite{Krishna2024}, and a few adders along with an 8-bit carry save adder.

\section{Application: Convolution-Based Edge Detection Using the Proposed Multiplier}
\label{sec:edge_detection}

To demonstrate the effectiveness of the proposed approximate signed multiplier in practical image processing, we applied it to an edge detection task using 2D spatial convolution. Edge detection is a fundamental operation in computer vision and serves as an ideal benchmark for evaluating arithmetic precision in approximate computing.
\begin{equation}
\text{Kernel} =
\begin{bmatrix}
-1 & -1 & -1 \\
-1 & \phantom{-}8 & -1 \\
-1 & -1 & -1 \\
\end{bmatrix}
\label{eq:laplacian_kernel}
\end{equation}

\begin{figure}[ht]
    \centering
    \includegraphics[scale=0.48]{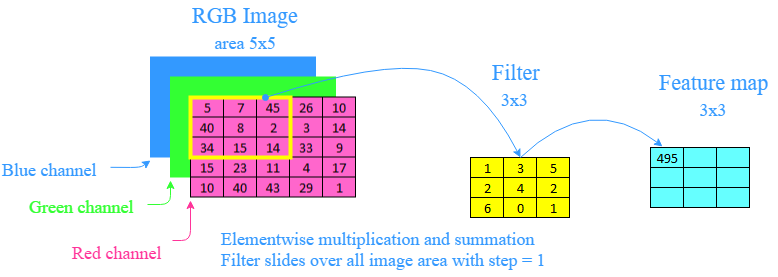}
    \caption{Illustration of the convolution process using the Laplacian kernel for edge detection on a $3 \times 3$ image patch..}
    \label{fig:laplacian_example}
\end{figure}

In this work, the standard multiplication operation in convolution was replaced with a custom element-wise multiplication based on the proposed multiplier. To perform this operation, a Python-based implementation was developed. Zero-padding was applied to the input image to preserve boundary pixels during filtering.
We employed the Laplacian kernel as shown in Equation~(\ref{eq:laplacian_kernel}), which emphasizes intensity discontinuities by computing the second spatial derivative of the image. Figure~\ref{fig:laplacian_example} shows the element-wise multiplication between the kernel and the input patch, followed by accumulation to produce the output pixel value.

\begin{figure}[ht]
    \centering
    \includegraphics[scale=0.11]{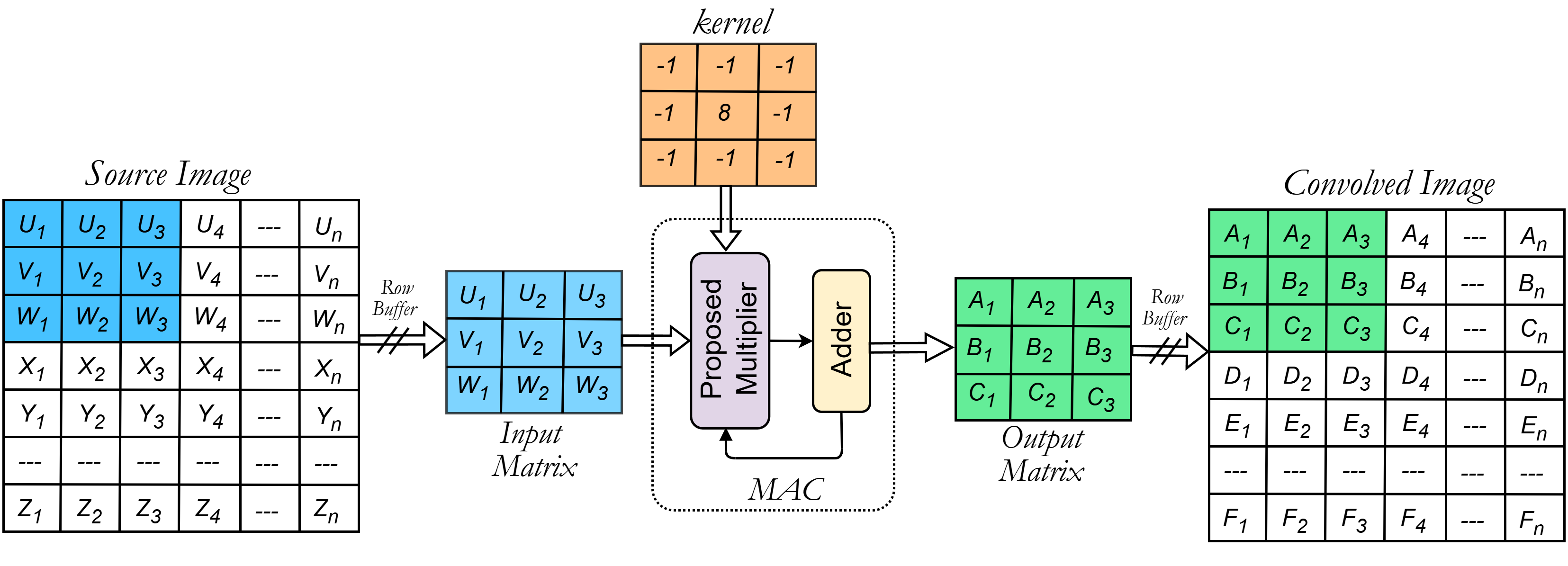}
    \caption{Hardware-oriented convolution framework using the proposed signed multiplier for edge detection.}
    \label{fig:conv_pipeline}
\end{figure}

Figure~\ref{fig:conv_pipeline} illustrates a hardware-efficient convolution framework tailored for edge detection on FPGA platforms. The architecture employs a row buffer to extract \(3 \times 3\) input patches, which are processed by a Multiply-Accumulate (MAC) unit using the proposed signed multiplier and a fixed Laplacian kernel. The accumulated results are stored in an output matrix to form the convolved image. Designed for streaming data, the system efficiently utilizes FPGA resources such as line buffers and DSP slices, enabling real-time edge detection with low power consumption.

\begin{figure*}[ht]
    \centering
    \begin{tabular}{c c c }
        \includegraphics[scale=0.283]{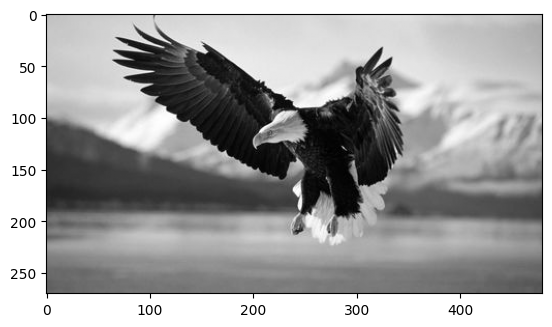} &
        \includegraphics[scale=0.283]{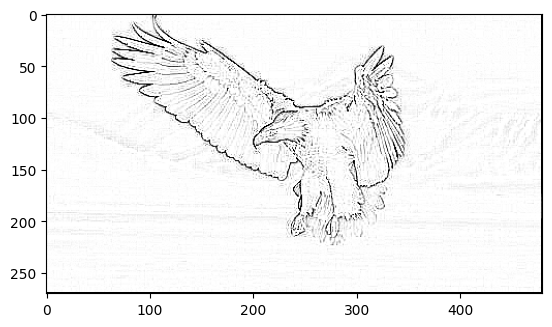} & 
        \includegraphics[scale=0.283]{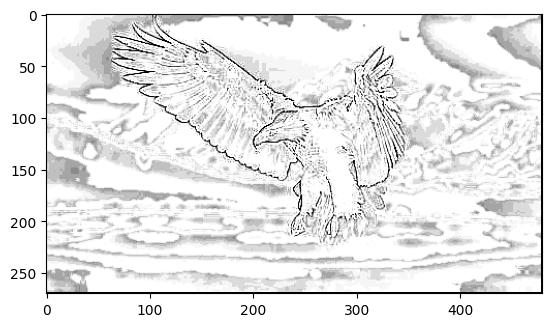} \\
        \scriptsize{Input Image} & \scriptsize{Exact Multiplier} & \scriptsize{Design~\cite{Esposito2018} (5.86 dB)} \\

        \includegraphics[scale=0.283]{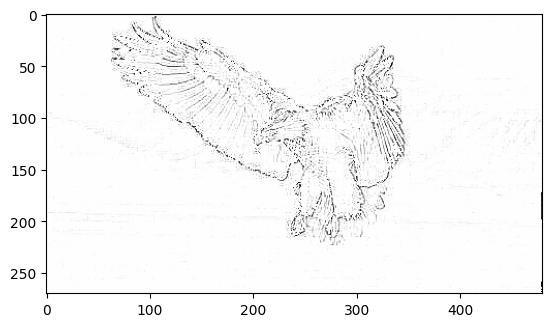} &
        \includegraphics[scale=0.283]{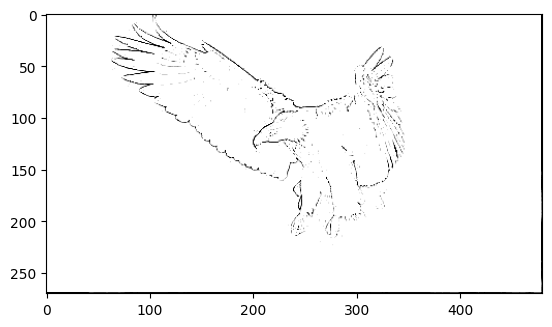} & 
        \includegraphics[scale=0.283]{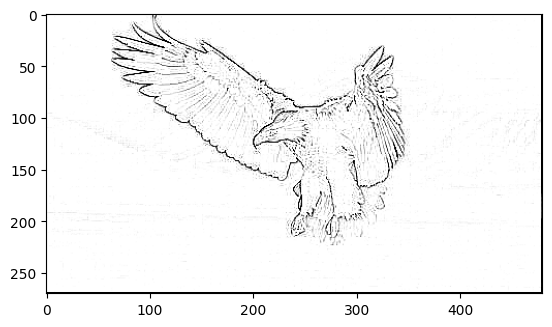} \\
        \scriptsize{Design~\cite{Guo2019} (10.84 dB)} & \scriptsize{Design~\cite{Du2022} (14.79 dB)} & \scriptsize{Proposed (20.13 dB)} \\
    \end{tabular}
    \caption{Edge detection outputs using the proposed and existing signed multipliers.}
    \label{fig:edge_images}
\end{figure*}

Figure~\ref{fig:edge_images} presents the output images resulting from edge detection using the proposed and several existing signed multipliers. The visual comparisons demonstrate how closely each approximate design preserves the edge structures relative to the exact multiplier output. To quantify the visual fidelity, Peak Signal-to-Noise Ratio (PSNR) values are reported for each design, computed with respect to the edge map generated by the exact multiplier. Higher PSNR values indicate better similarity to the exact reference, with the proposed design achieving the highest PSNR of 20.13~dB, demonstrating its superior accuracy.

\section{Results and Discussion}
\label{sec:results}

The edge detection application was implemented in 
Python to demonstrate the real-world applicability of the proposed design.
\subsection{Error Analysis}
The Mean Relative Error Distance (MRED) and Normalized Mean Error Distance (NMED) 
are widely used metrics to evaluate the accuracy of approximate arithmetic circuits. 
MRED quantifies the average of relative error distances across all test cases and is defined in Equation (\ref{eq:MRED})
\begin{equation}
\text{MRED} = \frac{1}{N} \sum_{i=1}^{N} \left| \frac{\text{Exact}_i - \text{Approx}_i}{\text{Exact}_i} \right|
\label{eq:MRED}
\end{equation}
NMED normalizes the total error with respect to the maximum exact output value and is given by Equation (\ref{eq:NMED})
\begin{equation}
\text{NMED} = \frac{1}{N} \sum_{i=1}^{N} \left( \frac{|\text{Exact}_i - \text{Approx}_i|}{\max(\text{Exact})} \right)
\label{eq:NMED}
\end{equation}
Table~\ref{tab:error_metrics} presents the error metrics of the proposed multiplier in comparison with existing designs. In this evaluation, various existing approximate compressor architectures were integrated into the proposed signed multiplier framework, and two key error metrics, such as NMED and MRED, were computed.

\begin{table}[h!]
    \centering
    \caption{Comparison of Error Metrics Between Proposed and Existing Multipliers}
    \label{tab:error_metrics}
    \begin{tabular}{|l|c|c|c|}
        \hline
        \textbf{Designs} & ER (\%) & NMED (\%) & MRED (\%) \\ 
        \hline
        Design~\cite{Strollo2020}      & 98.47 & 1.128 & 32.80  \\ 
        Design~\cite{Guo2019}          & 98.95 & 0.829 & 30.00  \\  
        Design~\cite{Esposito2018}     & 99.42 & 0.786 & 35.25 \\ 
        Design~\cite{Akbari2017}       & 97.37 & 0.738 & 29.02  \\ 
        Design~\cite{Krishna2024}      & 98.95 & 0.542 & 33.00  \\ 
        Design~\cite{Du2022}           & 98.15 & 0.731 & 26.84 \\ 
        \textbf{Proposed Design}       & 98.04 & 0.682 & 26.29 \\ 
        \hline
    \end{tabular}
\end{table}

\subsection{Hardware Analysis}
To evaluate the hardware-level characteristics of the proposed signed approximate multiplier, 
the design was described in Verilog HDL and synthesized using Synopsys Design Compiler with a UMC 90nm standard cell library under typical PVT (Process, Voltage, Temperature) conditions.
All the existing designs were evaluated in the same technology node as same as the proposed designs.
The proposed approximate multiplier achieves an energy improvement of $29.56\%$ compared to the best existing design reported in the recent work~\cite{Du2022} as shown in Table~\ref{comp}.

\begin{figure}[ht]
    \centering
    \includegraphics[scale=0.48]{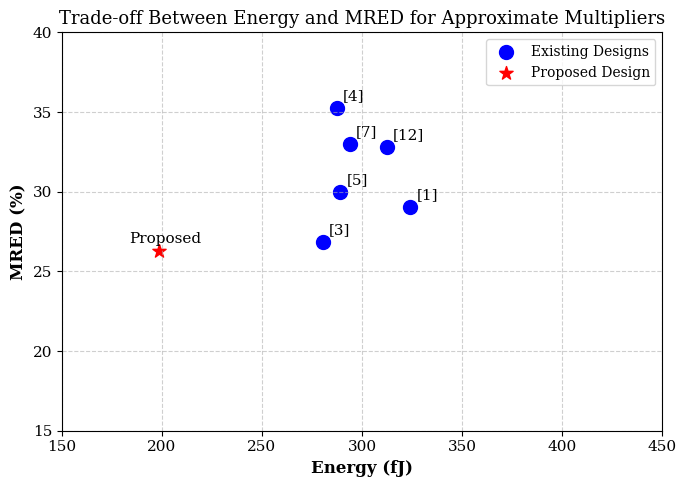} 
    \caption{Scatter plot showing the trade-off between energy and MRED for various approximate signed multipliers.}
    \label{fig:mred_pdp}
\end{figure}

Figure~\ref{fig:mred_pdp} presents a comparative scatter plot between Power-Delay Product (PDP) and MRED for several existing designs and the proposed approximate signed multiplier designs. Each point in the plot represents a design trade-off between arithmetic accuracy and energy efficiency. The proposed design, highlighted in red star, achieves both the lowest PDP and MRED, indicating a superior balance between performance and precision.

\begin{table}[ht]
\caption{Energy Comparison for Existing and Proposed  Multiplier Designs}
\label{comp}
\centering
    \begin{tabular}{|c|c|c|c|c|}
        \hline
        {\bf Design} & {\bf Area} ($\mu m^2$) & {\bf Power} (uW) & {\bf Delay} (ns) & {\bf PDP} (fJ) \\
        \hline
        Exact                          & 2204.75  & 178.10 & 3.28  & 584.17 \\ \hline
        Design \cite{Esposito2018}     & 1242.07  & 136.95 & 2.17  & 297.41 \\ \hline
        Design \cite{Akbari2017}       & 1972.91  & 122.19 & 2.65  & 324.08 \\ \hline
        Design \cite{Guo2019}          & 1164.34  & 116.05 & 2.49  & 289.15 \\ \hline
        Design \cite{Strollo2020}      & 1386.62  & 129.96 & 2.32  & 302.48 \\ \hline
        Design \cite{Krishna2024}      & 1306.84  & 124.89 & 2.35  & 293.95 \\ \hline
        Design \cite{Du2022}           & 1013.07  & 110.42 & 2.54  & 280.48 \\ \hline
        {\bf Proposed}                 & 809.23   & 94.52  & 2.10  & 198.54 \\ \hline
    \end{tabular}
\end{table}

\section{Conclusion}
\label{sec:conclusion}
This paper presents both accurate and approximate signed focus compressors designed for use in an approximate Baugh-Wooley signed multiplier. These compressors facilitate the efficient accumulation of both positive and negative partial products using a constant value. The proposed multiplier leverages truncation and an error compensation technique in the LSB portion to reduce computational errors while significantly lowering power consumption. Experimental results demonstrate that the proposed design saves 29\% of the energy compared to the best-performing existing design. Furthermore, its applicability to image processing tasks, such as edge detection, was validated, where it successfully preserved prominent edges with high accuracy.

\bibliographystyle{splncs04}
\bibliography{IEEEabrv,Binary.bib}

\begin{thebibliography}{10}
\providecommand{\url}[1]{\texttt{#1}}
\providecommand{\urlprefix}{URL }
\providecommand{\doi}[1]{https://doi.org/#1}

\bibitem{Akbari2017}
Akbari, O., Kamal, M., Afzali-Kusha, A., Pedram, M.: Dual-quality 4:2 compressors for utilizing in dynamic accuracy configurable multipliers. IEEE Transactions on Very Large Scale Integration (VLSI) Systems  \textbf{25}(4),  1352--1361 (2017). \doi{10.1109/TVLSI.2016.2643003}

\bibitem{Du2022}
Du, L., Ni, L., Liu, X., Mao, W., Yu, H.: A low-power approximate multiplier with sign-focus compressor and error compensation. In: 2022 IEEE Asia Pacific Conference on Circuits and Systems (APCCAS). pp. 226--230 (2022). \doi{10.1109/APCCAS55924.2022.10090316}

\bibitem{Laimin}
Du, L., Ni, L., Liu, X., Peng, G., Li, K., Mao, W., Yu, H.: A low-power dnn accelerator with mean-error-minimized approximate signed multiplier. IEEE Open Journal of Circuits and Systems  \textbf{5},  57--68 (2024). \doi{10.1109/OJCAS.2023.3279251}

\bibitem{Esposito2018}
Esposito, D., Strollo, A.G.M., Napoli, E., De~Caro, D., Petra, N.: Approximate multipliers based on new approximate compressors. IEEE Transactions on Circuits and Systems I: Regular Papers  \textbf{65}(12),  4169--4182 (2018). \doi{10.1109/TCSI.2018.2839266}

\bibitem{Guo2019}
Guo, Y., Sun, H., Kimura, S.: Energy-efficient and high-speed approximate signed multipliers with sign-focused compressors. In: 2019 32nd IEEE International System-on-Chip Conference (SOCC). pp. 330--335 (2019). \doi{10.1109/SOCC46988.2019.1570548436}

\bibitem{han2013approximate}
Han, J., Orshansky, M.: Approximate computing: An emerging paradigm for energy-efficient design. IEEE European Test Symposium (ETS) pp.~1--6 (2013)

\bibitem{Krishna2024}
Krishna, L.H., Sk, A., Rao, J.B., Veeramachaneni, S., Sk, N.M.: Energy-efficient approximate multiplier design with lesser error rate using the probability-based approximate 4:2 compressor. IEEE Embedded Systems Letters  \textbf{16}(2),  134--137 (2024). \doi{10.1109/LES.2023.3280199}

\bibitem{hemanth2025}
Krishna, L.H., Veeramachaneni, S., Bodapati, S., Jammu, B., Sk, N.M.: Optimizing multipliers: An energy-efficient design using a novel 3:2 compressor. In: 2025 38th International Conference on VLSI Design and 2024 23rd International Conference on Embedded Systems (VLSID). pp. 127--132 (2025). \doi{10.1109/VLSID64188.2025.00035}

\bibitem{baugh1974}
Kroft, D.: Comments on "a two's complement parallel array multiplication algorithm". IEEE Transactions on Computers  \textbf{C-23}(12),  1327--1328 (1974). \doi{10.1109/T-C.1974.223863}

\bibitem{mittal2016survey}
Mittal, S.: A survey of techniques for approximate computing. ACM Computing Surveys (CSUR)  \textbf{48}(4),  1--33 (2016)

\bibitem{Booth1975}
Rubinfield, L.: A proof of the modified booth's algorithm for multiplication. IEEE Transactions on Computers  \textbf{C-24}(10),  1014--1015 (1975). \doi{10.1109/T-C.1975.224114}

\bibitem{Strollo2020}
Strollo, A.G.M., Napoli, E., De~Caro, D., Petra, N., Meo, G.D.: Comparison and extension of approximate 4-2 compressors for low-power approximate multipliers. IEEE Transactions on Circuits and Systems I: Regular Papers  \textbf{67}(9),  3021--3034 (2020). \doi{10.1109/TCSI.2020.2988353}

\bibitem{venkataramani2015approximate}
Venkataramani, S., Raghunathan, A.: Approximate computing and the quest for computing efficiency. Proceedings of the 52nd Annual Design Automation Conference (DAC) pp.~1--6 (2015)

\end{thebibliography}

\end{document}